\documentclass[final]{IEEEtran}
\newlength \figwidth
\usepackage{cite}
\ifCLASSINFOpdf
  \usepackage[pdftex]{graphicx}
	\usepackage{epstopdf}
  \graphicspath{{../Figures/}}
  \DeclareGraphicsExtensions{.eps}
\else
  \usepackage[dvips,draft]{graphicx}
\fi
\usepackage[cmex10]{amsmath}
\usepackage{graphicx}
\usepackage[caption=false]{subfig}
\usepackage{amssymb}
\usepackage{pifont}
\usepackage{amsthm}
\usepackage{url}
\usepackage{dsfont}
\usepackage{tabulary}
\usepackage{multirow}
\usepackage{color}

\usepackage{color}
\usepackage{times}
\usepackage{verbatim}
\usepackage{floatflt}
\usepackage{enumerate}
\usepackage{array}
\usepackage{multicol,afterpage,wrapfig} 
\usepackage{tikz}
\usepackage[noend]{algpseudocode}
\makeatletter
\def\BState{\State\hskip-\ALG@thistlm}
\makeatother
\usepackage{amsmath,amssymb,eucal}
\usepackage[utf8]{inputenc}
\usepackage{epsfig}
\usepackage{exscale}
\usepackage{latexsym}
\usepackage{verbatim}
\usepackage{amsfonts}
\usepackage{multirow}
\usepackage{cite}
\usepackage{balance}
\usepackage{color}
\usepackage{transparent}
\usepackage{import}
\usepackage{mathtools}
\usepackage{tikz}
\usetikzlibrary{automata,arrows,positioning,calc}
\usepackage{bbm}
\usepackage[printonlyused,withpage]{acronym}
\usepackage{tabu}
\usepackage{longtable}
\usepackage{balance}
\usepackage{footnote}
\usepackage{tablefootnote}
\usepackage{enumitem}

\usepackage{adjustbox}
\usepackage{multirow}
\usepackage{color, colortbl}
 	\definecolor{lightsalmon}{rgb}{1.0, 0.63, 0.48}
\usepackage{pifont}
\usepackage[normalem]{ulem}

\def\BibTeX{{\rm B\kern-.05em{\sc i\kern-.025em b}\kern-.08em
    T\kern-.1667em\lower.7ex\hbox{E}\kern-.125emX}}
    
\renewcommand{\IEEEQED}{\IEEEQEDopen}

\newcommand*\xbar[1]{%
  \hbox{%
    \vbox{%
      \hrule height 0.5pt 
      \kern0.36ex
      \hbox{%
        \kern-0.12em
        \ensuremath{#1}%
        \kern-0.12em
      }%
    }%
  }%
}

\newfont{\bbb}{msbm10 scaled 500}

\newfont{\bb}{msbm10 scaled 1100}

\newcommand{\executeiffilenewer}[3]{%
\ifnum\pdfstrcmp{\pdffilemoddate{#1}}%
{\pdffilemoddate{#2}}>0%
{\immediate\write18{#3}}\fi%
}

\usepackage[
top    = 0.7in,
bottom = 0.95in,
left   = 0.66in,
right  = 0.68in]{geometry}

\usetikzlibrary{arrows,calc}
\allowdisplaybreaks

\IEEEoverridecommandlockouts
\begin{document}
\pagenumbering{gobble}

\newtheorem{Theorem}{\bf Theorem}
\newtheorem{Corollary}{\bf Corollary}
\newtheorem{Remark}{\bf Remark}
\newtheorem{Lemma}{\bf Lemma}
\newtheorem{Proposition}{\bf Proposition}
\newtheorem{Assumption}{\bf Assumption}
\newtheorem{Definition}{\bf Definition}

\title{On the Latency of IEEE 802.11ax WLANs with Parameterized Spatial Reuse
\thanks{This work was supported by the European Union’s Horizon 2020 research and innovation programme under Marie Skłodowska-Curie Grant agreement No.812991 (PAINLESS). The work of G.~Geraci was supported in part by MINECO under Project RTI2018-101040-A-I00 and by the Postdoctoral Junior Leader Fellowship Programme from ``la Caixa" Banking Foundation.}}

\author{\IEEEauthorblockN{{Eloise~de~Carvalho~Rodrigues$^{\star \dagger}$, Adrian~Garcia-Rodriguez$^{\star}$, Lorenzo~Galati~Giordano$^{\star}$, and Giovanni~Geraci$^{\dagger}$}}\\ \vspace{0.2cm}
\IEEEauthorblockA{$^{\star}$\emph{Nokia Bell Labs, Dublin, Ireland}\\
\IEEEauthorblockA{$^{\dagger}$\emph{Universitat Pompeu Fabra (UPF), Barcelona, Spain}
}}}

\maketitle

\thispagestyle{empty}
\begin{abstract}
In this article, we evaluate the performance of the parameterized spatial reuse (PSR) framework of IEEE 802.11ax, mainly focusing on its impact on transmission latency. Based on detailed standard-compliant system-level simulations, we provide a realistic analysis of the effects of PSR considering different scenario densities, traffic loads, and access points (APs) antenna capabilities to quantify its performance gains under various scenarios.  Our results show that, in medium-density scenarios, PSR can offer up to a $3.8\times$ reduction in the 5\%  worst-case latencies for delay-sensitive stations with respect to an 802.11ax system without PSR. Moreover, our study demonstrates that, for low-latency communications, providing the network with PSR capabilities may be an appealing alternative to the deployment of more costly multi-antenna APs.
\end{abstract}
\IEEEpeerreviewmaketitle

 \section{Introduction}


From our houses to smart cities and factories, Wi-Fi products are widely used due to the simplicity of the IEEE 802.11 protocols combined with low-cost deployment and management. 
From 2020 to 2023, the number of Wi-Fi 6 hotspots---based on IEEE 802.11ax---is foreseen to increase 13 fold \cite{cisco_2020}. As the number of concurrently connected devices increases, one of the main challenges to overcome is the potential delay introduced by the mandatory listen-before-talk (LBT) adopted by technologies using the unlicensed spectrum \cite{deng2017ieee}. LBT imposes that a device willing to access the channel must listen to the medium and guarantee that there are no other ongoing transmissions before attempting to transmit. This mechanism is a potential show stopper for low-latency applications, e.g., Industry 4.0 and autonomous robots.

A way to address this problem is by increasing spatial reuse (SR). Two related techniques are included in the IEEE 802.11ax amendment: overlapping basic service set packet detect (OBSS/PD) and parameterized spatial reuse (PSR)\footnote{Referred to as spatial reuse parameter (SRP) until Standard Draft 5.0.} \cite{ieee}. OBSS/PD consists of relaxing the clear channel assessment (CCA) signal detection threshold to increase the probability of concurrent transmissions. Instead, PSR takes advantage of the trigger frame (TF) that precedes uplink (UL) transmissions to permit access points (APs) to control spatial reuse. According to the information conveyed in the TF, non-associated stations (STAs) determine whether they are allowed to access the channel based on the potential interference they may generate towards the AP that transmitted the TF. This mode of operation can be advantageous for critical applications with low-latency requirements. Indeed, IEEE 802.11be---the basis of the next-generation Wi-Fi standard---is currently discussing how to further reduce worst-case latencies by building on top of the PSR framework \cite{psrContribution, lopez2019ieee, KhoLevAky2020, garciarodriguez2020ieee}. In these circumstances, assessing the gains already provided by the PSR scheme available in IEEE 802.11ax is of crucial importance.

 Related works surveying the main functionalities and operation of both OBSS/PD and PSR spatial reuse include \cite{qu2019survey,  khorov2018tutorial, afaqui2016ieee, wilhelmi2019spatial, valkanis2019ieee}, while simulation-based performance evaluations of the OBSS/PD SR technique are provided in \cite{shen2018research, wilhelmi2019performance, wilhelmi2019spatial}. To the best of our knowledge, there are currently no works on the literature characterizing PSR's performance. This is mainly due to the complexity of implementing and evaluating the PSR functionality since an accurate analysis also requires the implementation of intricate 802.11ax features such as TF-based operations and UL multi-user multiple-input and multiple-output (MU-MIMO) \cite{wilhelmi2019spatial}. 


In this paper, we compare the performance of an IEEE 802.11ax system with and without PSR capabilities. In particular, we analyze how PSR affects latency in practical scenarios with various deployment densities, traffic loads, and number of AP antennas. To do so, we implement the PSR framework as specified by IEEE 802.11ax, accounting for its complex features and the latest 3GPP 3D channel model, all leading to accurate results. Our main takeaways can be summarized as follows:
\begin{itemize}[leftmargin=*]

\item For high-, medium-, and low-density deployments with high traffic loads, PSR can reduce the 5\% worst-case latencies with respect to the baseline IEEE 802.11ax operations by a factor 1.8, 3.8, and 3.4, respectively.
\item Allowing a more aggressive spatial reuse through PSR provides more remarkable latency benefits under higher traffic loads---where the wireless medium remains generally occupied---than under low-to-medium loads.
\item Latency-wise, a PSR system with single-antenna APs can outperform eight-antenna APs without spatial reuse ca\-pa\-bi\-li\-ties, designating PSR as an alternative to more costly multi-antenna APs.
\end{itemize}
\section{PSR-based Spatial Reuse Operation}

In this section, we describe both baseline (Sec. \ref{sec:baseline}) and PSR-enabled (Sec. \ref{sec:psrop}) IEEE 802.11ax operations, providing examples to highlight the relevance of PSR for low-latency applications. 
\label{sec:psr}


\subsection{Baseline 802.11ax Operation}
\label{sec:baseline}

In the baseline 802.11ax operation, the transmission is done through carrier sense multiple access/collision avoidance (CSMA/CA). This entails that APs and STAs need to perform i) physical and ii) virtual carrier sensing before initiating a transmission. Physical carrier sensing determines whether the channel is occupied or not by listening to wireless signal energy. Virtual carrier sense relies on the transmitter informing the surrounding nodes about the length of the imminent data transmissions. The STAs receiving this signaling set their network allocation vector (NAV) that indicates for how long they must defer from accessing the medium. The medium is sensed as ``busy'' if the STA/AP  either i) senses signals at or above -62 dBm, and/or ii) the NAV is set \cite{perahia2013next}. Once an AP gains channel access in TF-based operations, it transmits a TF to initiate an uplink transmission of one or multiple STAs. The TF is a control frame containing information about the scheduled stations and other relevant information.

\begin{figure}[!t]

\subfloat[Baseline 802.11ax scenario]{%
  \includegraphics[clip,width=\columnwidth]{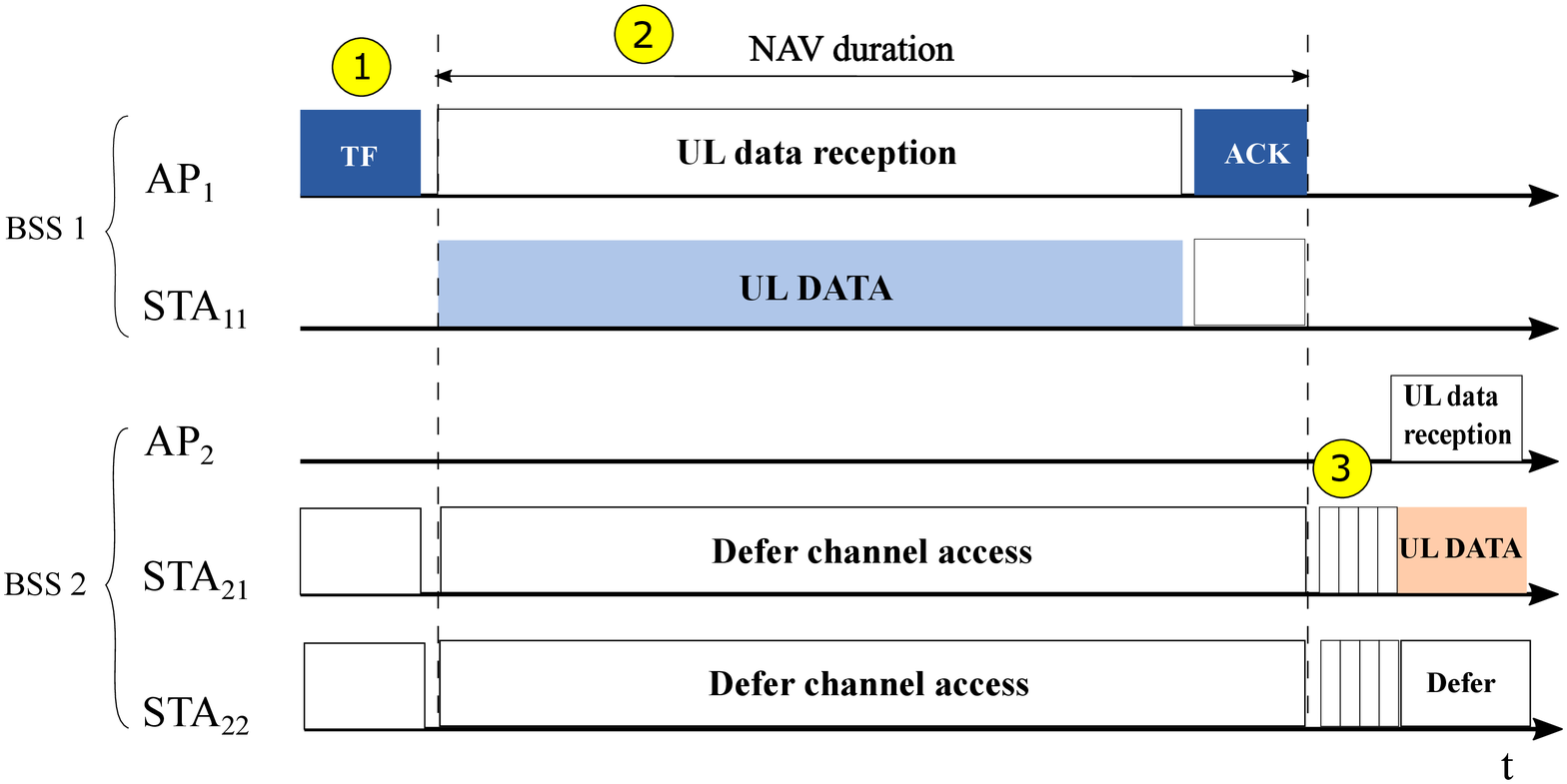}%
}

\subfloat[PSR-enabled scenario]{%
  \includegraphics[clip,width=0.95\columnwidth]{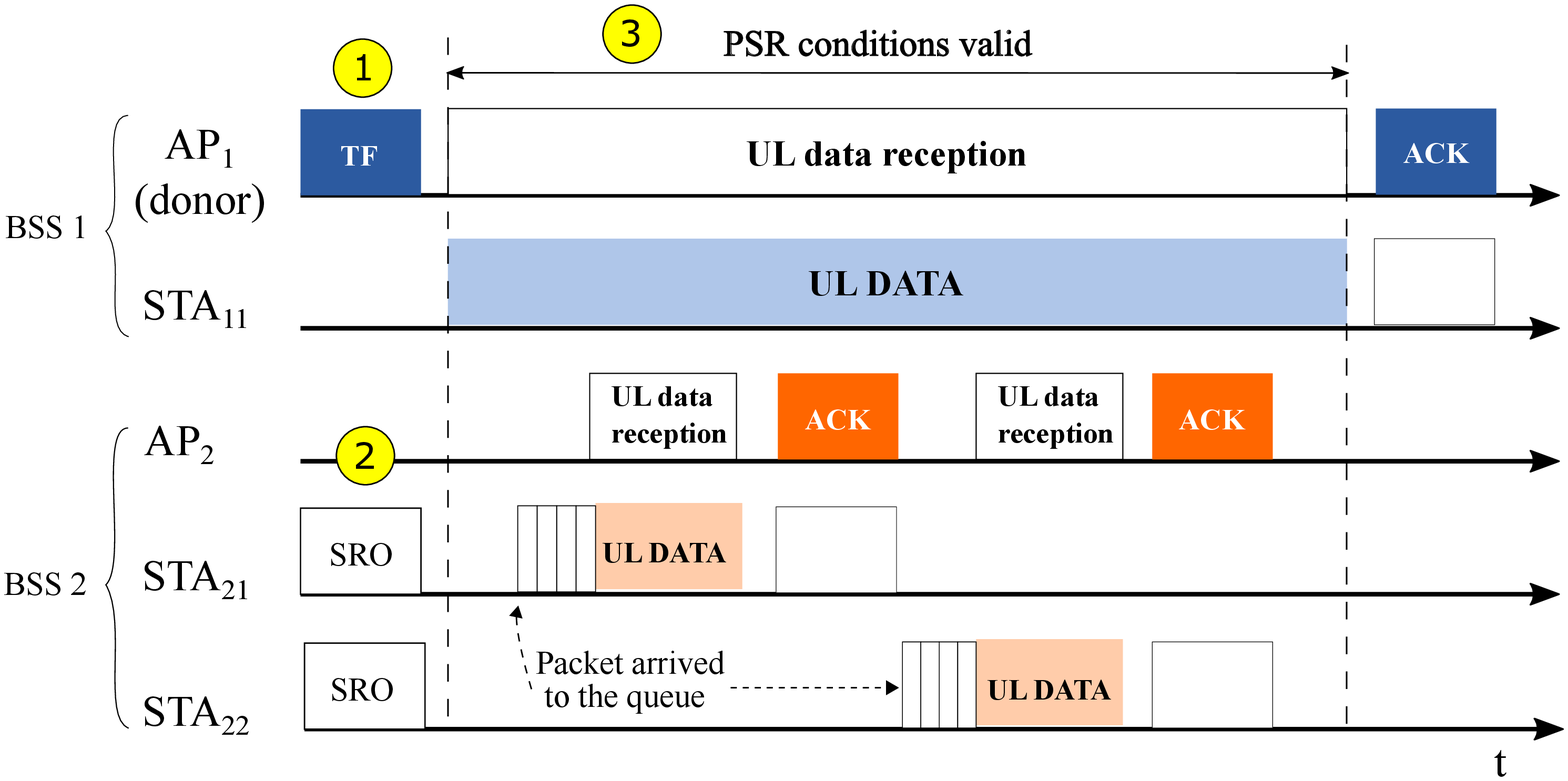}%
}

\caption{Illustration of (a) baseline operation for an IEEE 802.11ax system using CSMA/CA, and (b) a PSR-capable IEEE 802.11ax system.}
\label{fig:psr}

\end{figure}

Fig. \ref{fig:psr}(a) illustrates legacy 802.11 operations. In the example, we have two basic service sets (BSSs): BSS 1, composed by AP$_1$ and STA$_{11}$, and BSS 2, composed by AP$_2$, STA$_{21}$ and STA$_{22}$. The STAs from BSS 2 are latency-sensitive stations, meaning that their packets are short and must be delivered in a timely manner. In this example, AP$_1$ is going to schedule a long uplink transmission from STA$_{11}$ (e.g., in the order of several milliseconds). In Fig. \ref{fig:psr}(a), once AP$_1$ gains channel access in accordance with the aforementioned rules, the operation continues as follows:
\begin{enumerate}[leftmargin=*]
    \item AP$_1$ triggers an uplink transmission from STA$_{11}$;
    \item STA$_{21}$ and STA$_{22}$ set their NAV indicating that the channel is busy and defer from channel access until the NAV expires;
    \item STA$_{21}$ and STA$_{22}$ contend for channel access again. STA$_{21}$ gets channel access first, and STA$_{22}$ will defer from channel access.
\end{enumerate}{}
In this example, the latency-sensitive STAs (STA$_{21}$ and STA$_{22}$) may suffer for having to wait for the long transmission of STA$_{11}$ to be finished before accessing the channel and transmitting their short packets, which have a negative impact on their perceived quality of service.

\subsection{PSR Operation}
\label{sec:psrop}

 The PSR-based spatial reuse mechanism was included in the IEEE 802.11ax standard to allow the wireless medium to be reused more often in dense deployment scenarios \cite{ieee, qu2019survey}. Compared to other SR techniques such as OBSS/PD, PSR facilitates a more controlled spatial reuse. This control is achieved because APs, the \textit{transmission holders} or \textit{donors}, only provide to non-associated STAs the possibility of ac\-cess\-ing the channel during the uplink-triggered transmissions as long as they comply with a specified acceptable level of interference, devised so as not to harm the uplink of own served STAs. In other words, the donor APs that want to facilitate spatial reuse do so in such a way that the performance of its scheduled uplink transmissions is guaranteed. For this, the AP defines an \textit{acceptable interference  level}, $\text{I}_{\text{AP}}$, as the maximum interference that can be perceived by the transmission holder without compromising its trans\-mis\-sion. This acceptable interference is a value in dBm calculated as: 
 \begin{equation}
     \text{I}_{\text{AP}} = \text{UL\_Target\_RSSI} - \text{Min\_SNR\_MCS} - \text{Safety\_Margin},
\end{equation} 
where $\text{UL\_Target\_RSSI}$  is the UL target receive signal strength indicator (RSSI) in dBm indicated in the TF, $\text{Min\_SNR\_MCS}$ is the minimum signal to noise ratio (SNR) value that yields $\leq 10\%$ packet error rate (PER) for the highest modulation and coding scheme (MCS) of the following uplink data trans\-mis\-sion, and a $\text{Safety\_Margin}$ that should not exceed 5 dB.
 
To make sure that the interference received is below $\text{I}_{\text{AP}}$, the AP includes the $\text{PSR\_INPUT}$ field in the TF. The value of  $\text{PSR\_INPUT}$ is computed as:
\begin{equation}
\label{PSRinput}
    \text{PSR\_INPUT} = \text{TX\_PWR}_{\text{AP}} + \text{I}_{\text{AP}},
\end{equation}
 where $\text{TX\_PWR}_{\text{AP}}$ is the power used by the AP to transmit the trigger frame in dBm.

A STA with PSR capabilities that receives a TF allowing PSR is allowed to resume the channel access contention procedure and initiate a transmission after identifying what is called a \textit{spatial reuse opportunity} (SRO) if the following conditions, designed to avoid interference in reception for the AP that holds the current transmission, are met:
\subsubsection*{Condition 1}
The TF received by the STA is an inter-BSS trigger frame, i.e. a trigger frame sent from an AP located in a BSS different from the BSS in which the STA willing to use SR is associated;
\subsubsection*{Condition 2}
The intended transmission power $\text{TX\_PWR}_{\text{STA}}$ should not exceed the interference limit set by the AP granting the spatial reuse opportunity. This condition is formally defined as:
    \begin{equation}
\label{intendedTXpower}
    \text{TX\_PWR}_{\text{STA}} - 10\log_{10}\left({\frac{\text{TX\_BW}_{\text{STA}}}{20 \text{MHz}}}\right) \leq  \text{PSR\_INPUT} - \text{RPL},
\end{equation}
where $\text{TX\_BW}_{\text{STA}}$ is the intended transmission bandwidth in MHz and $\text{RPL}$ is the received power level that indicates the receiving power of TF \cite{qu2019survey}.

Importantly, the conditions set by the AP allowing PSR are only valid for the duration of the uplink data transmission that follows the TF. This entails that the STA intending to transmit during an SR opportunity should make sure that its transmission duration is smaller than the uplink-triggered transmission initiated by the donor AP. Once the STA decides to take the PSR opportunity, it may ignore the network allocation vector (NAV), but it must still perform physical carrier sensing.

To summarize the process, consider the example in Fig. \ref{fig:psr}(b) in which the APs and the latency-sensitive STAs now have spatial reuse capabilities. The PSR operation, in this example, would proceed as follows:
\begin{enumerate} [leftmargin=*]
    \item The donor AP$_1$ triggers STA$_{11}$ for uplink transmission and sets the trigger frame field that allows PSR, also setting $\text{PSR\_INPUT}$;
    \item STA$_{21}$ and STA$_{22}$ receive an inter-BSS trigger frame allowing PSR. Both stations can grab spatial reuse opportunities as we assume that Conditions 1 and 2 are met;
    \item As the transmission triggered by AP$_1$ is very long, both STA$_{21}$ and STA$_{22}$ are able to transmit their short packets within the currently set spatial reuse opportunity as soon as the packets arrive in the queue. 

\end{enumerate}{}

This example clearly illustrates how PSR may help to reduce the worst-case latencies of the STAs with latency constraints by taking advantage of the multiple short-packet transmissions that may be performed within a long uplink-triggered transmission.

\section{PSR Performance Evaluation}
\label{sec:Performance}

In this section, we consider the scenario deployment described in Sec. \ref{sec:SystemModel} to extensively evaluate the impact of deployment densities (Sec. \ref{4a}), traffic loads (Sec. \ref{4b}), and the number of antennas at the APs (Sec. \ref{4c}) on the performance of two 802.11ax systems: a baseline one without PSR capabilities, and an upgraded one where devices have PSR capabilities. To evaluate the performance of each scenario, we run standard-compliant system-level simulations that account for regulatory constraints, accurate channel and traffic modeling and the main IEEE 802.11ax features.

\subsection{System Configuration}
\label{sec:SystemModel}
\begin{figure}[!t]
\centering
\includegraphics[width=\columnwidth]{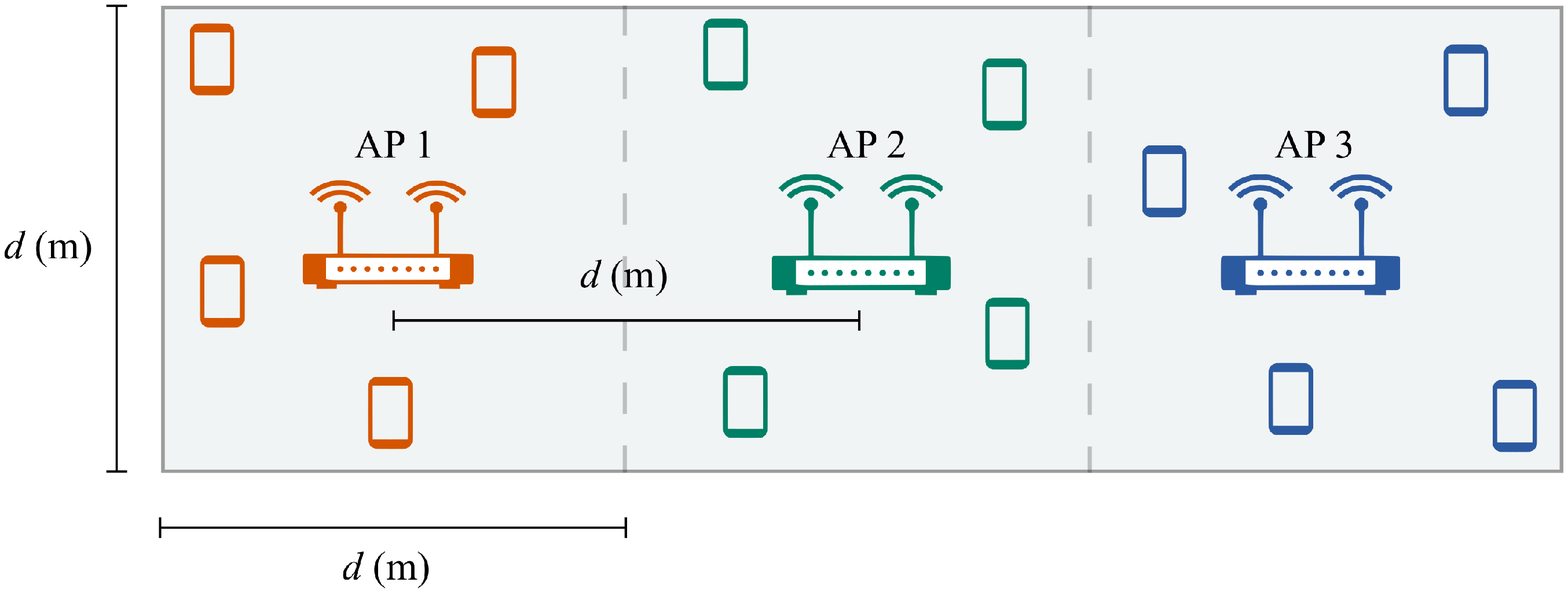}
\caption{Illustration of a $3d \times d$ indoor scenario with three ceiling-mounted Wi-Fi APs separated by a distance $d$ and $s$ STAs uniformly distributed.}

\label{fig:scenario}
\end{figure}

The deployed scenario, illustrated in Fig. \ref{fig:scenario}, consists in a $3d \times d$ single-floor indoor wireless local area network (WLAN).  This scenario comprises three ceiling-mounted Wi-Fi APs at the height of 3 m separated by a distance $d$ and 24 STAs uniformly deployed at the height of 1 m in the considered area. The stations associate to the APs based on the strongest average received signal strength (RSS). We assume that, initially, all the APs are equipped with four antennas, and all the STAs are equipped with one omnidirectional antenna. As PSR operates in uplink, only uplink transmissions are considered.
\begin{table}
\centering
\caption{System parameters}
\label{table:parameters}
\def\arraystretch{1.1}
\begin{tabulary}{0.75\columnwidth}{ |p{2.9cm} | p{5.1cm} | }
\hline
\rowcolor{lightsalmon}
\textbf{Parameter}    & Description\\ \hline

  \textbf{AP deployment}    &  3 ceiling mounted APs; AP height = 3 m
 \\ \hline
    \textbf{STA deployment} & 16 broadband STAs, 8 low-latency STAs; 2D uniform distribution; STA height = 1 m
 \\ \hline
     \textbf{STA-AP association} & Strongest average received signal
 \\ \hline
     \textbf{Frequency/Bandwidth}    & 5.18GHz/80MHz (1 channel) \\ \hline 
     \textbf{AP/STA max. TX power}    & 24 dBm/15 dBm\\ \hline 
     \textbf{AP/STA noise figure}    & 7 dB/ 9 dB\\ \hline 
     \textbf{Num. antennas per STA}    &  1 omni. antenna \\ \hline 
     \textbf{AP spatial Filter}    & Zero Forcing  \\ \hline 
     \textbf{AP/STA sensitivity}    &  -90 dBm  \\ \hline 
     \textbf{PSR Safety\_Margin} &  3 dB \\ \hline
     \textbf{MCS selection} & SINR-driven \\ \hline
    \textbf{STA scheduling} & Round-Robin scheduling policy \\ \hline
    \textbf{Max. TXOP length} & 4 ms \\ \hline
    \textbf{Broadband traffic} &  FTP3 \cite{broadband} with a file size of 0.5 MBytes \\ \hline
    \textbf{Low-latency traffic} &  Augmented reality \cite{lowlatency} with file size of 32 bytes and constant file arrival of 10 ms \\ \hline
    \textbf{Channel model} & 3D spatial channel model (3GPP TR38.901 – InH \cite{3gpp})\\ \hline
     
    \textbf{AP/STA MAC conf.} & IP/MAC header overhead considered; No EDCA; No RTS/CTS \\ \hline
    \textbf{AP/STA PHY conf.} & PHY header overhead considered; 11ax OFDM numerology with cyclic prefix duration = 0.8 $\mathrm{\mu}$s; Perfect channel state information acquisition
\\ \hline 
  \textbf{Simulations setup}    &  Monte Carlo simulations, 10 drops of 2 s for each scenario
 \\ \hline
    
\end{tabulary}
\end{table}

Out of the 24 STAs, 16 STAs have an uplink broadband traffic modeled as a File Transfer Protocol (FTP3) service \cite{broadband} with a file size of 0.5 MB and 8 STAs have low-latency traffic modeled as an augmented reality (AR) application \cite{lowlatency}, with a file size of 32 bytes and files arriving periodically at a frequency of 10 ms. For simplicity, in the rest of the paper, the STAs with FTP3 traffic are referred to as \textit{broadband traffic} stations, while the STAs with periodic file arrival traffic are referred to as \textit{low-latency traffic} stations. There is a fundamental challenge when it comes to maintaining acceptable performance in this mixed traffic scenario: as STAs share the same channel, broadband stations generate most of the load in the system and might be scheduled for long transmissions while the low-latency stations, that should meet latency deadlines, might continuously find the channel busy. 

In the baseline scenario, APs implement all the main IEEE 802.11ax features. When triggering uplink, APs spatially multiplex as many STAs as possible per transmission opportunity. For the PSR deployment, on top of the baseline functionalities, STAs can contend for channel access, but only when PSR-based channel access opportunities are identified. With the main objective of reducing the worst-case latencies of the low-latency traffic STAs, following the explanation introduced in the example given in Sec. \ref{sec:psr}, we assume that in the PSR scenario all APs enable PSR operations, and only low-latency traffic stations can grab those PSR opportunities.

Application files are generated by the traffic model and segmented into the transport layer, including the transport layer headers. The packet is then converted into internet protocol (IP) packets in the IP layer. The IP packets are then mapped into medium access control (MAC) layer frames and physical (PHY) layer symbols subsequently, with headers and overhead considered and variable PHY header sizes depending on transmission characteristics.

 The channel model is based on the 3D Indoor Office (InH) channel model from the latest 3GPP Technical Report 38.901  \cite{3gpp}, which accounts for 3D channel directionality in the large-scale and small-scale fading parameters. The algorithm used for MCS selection is a signal-to-interference-plus-noise-ratio (SINR)-driven approach, in which we assume that the transmitting STA knows what the active and transmitting nodes are. All the APs employ a Round Robin scheduler, with strict priority given to low-latency STAs.
 
 A summary of the system parameters is shown in Table \ref{table:parameters}.

\subsection{Deployment Density}
\label{4a}

In this section, we consider three different deployment densities: higher, medium, and lower density, with $d$ in the scenario described in Fig. \ref{fig:scenario} varying from 10 to 20 and 30 meters, respectively, and traffic load of 100  Mbps for broadband stations.  

\begin{figure}[!t]
\centering
\includegraphics[width=\columnwidth]{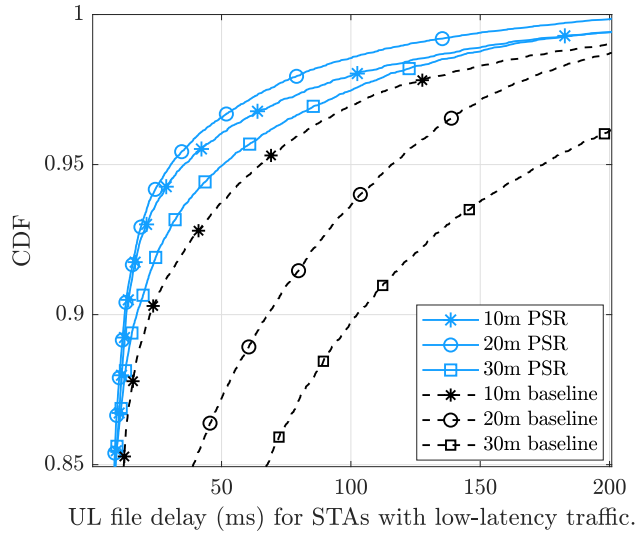}
\caption{CDF of the uplink file delay for low-latency STAs, inter-AP distances $d = \{ 10, 20, 30\}$m, and 100 Mbps broadband traffic load.} 

\label{fig:delay}
\end{figure}

Fig. \ref{fig:delay} illustrates the impact of PSR on the worst file delays experienced by the STAs generating low-latency traffic. Overall, we can observe that, for all densities considered, when the 802.11ax PSR feature is activated (solid curves), smaller delays can be achieved compared to baseline configurations without such feature. Indeed, when compared to the baseline 802.11ax system,  the PSR feature is capable of reducing the 5\%-worst delays by a factor of approximately $1.8\times, 3.8\times, 3.4\times$ for inter-AP distances of 10, 20, and 30 m, respectively.

\begin{figure}[!t]
\centering
\includegraphics[width=\columnwidth]{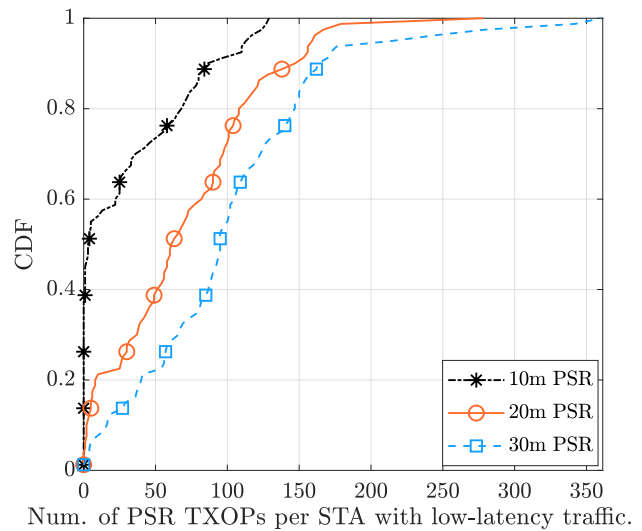}
\caption{CDF of the number of PSR-based transmissions gained per low-latency STA, inter-AP distances $d = \{ 10, 20, 30\}$m, and 100 Mbps broadband traffic load.}

\label{fig:srptxop}
\end{figure}

The following conclusions can be made from  Fig. \ref{fig:delay}:
\begin{itemize} [leftmargin=*]
    \item    For the baseline 802.11ax system, the worst-case delays are reduced as the deployment density increases. This is because the STAs are closer to their serving APs in denser scenarios, which allows the STAs to transmit using higher MCS levels and reduces their file transmission duration. In turn, this results in reduced channel access delays since the wireless channel becomes less occupied. Moreover, some devices might not be able to detect other ongoing transmissions in low-density scenarios due to the distance, which leads to a higher number of collisions and retransmissions \cite{perahia2013next}.

\item Instead, for the 802.11ax system with PSR capabilities, the worst-case file transmission delays are reduced when the inter-AP distance grows from 10 m to 20 m. This behavior can be explained by observing Fig.\ \ref{fig:srptxop}, which illustrates the CDF of the number of PSR-based gained transmissions per station for each of the deployment densities of Fig.\ \ref{fig:delay}. Fig. \ref{fig:srptxop} demonstrates that low-latency STAs have the lowest probability of gaining PSR-based transmissions in the densest deployment considered, with more than 40\% of the stations not being capable of finding SR channel access opportunities. This happens because in the densest deployment scenario they i) are less likely to find SROs---since the increased RPL means that they cannot satisfy the power constraint set by the AP granting the SR opportunity as per \eqref{intendedTXpower}---, and/or ii) deem the medium as busy when---due to uplink transmissions from STAs triggered by the donor AP---their RPL is higher than the physical carrier sensing threshold. 

\item Furthermore, the results of Fig.\ \ref{fig:delay} illustrate that the worst-case transmission delays under PSR increase when the inter-AP distances grow from 20 m to 30 m. This is because the performance loss resulting from having higher distances between STAs and their serving APs---following the same arguments as the corresponding baseline scenario---outweighs the gains introduced by the slightly higher number of SROs found, as per Fig.\ \ref{fig:srptxop}.

\end{itemize}{}

\begin{figure}[t]
\centering
\includegraphics[width=\columnwidth]{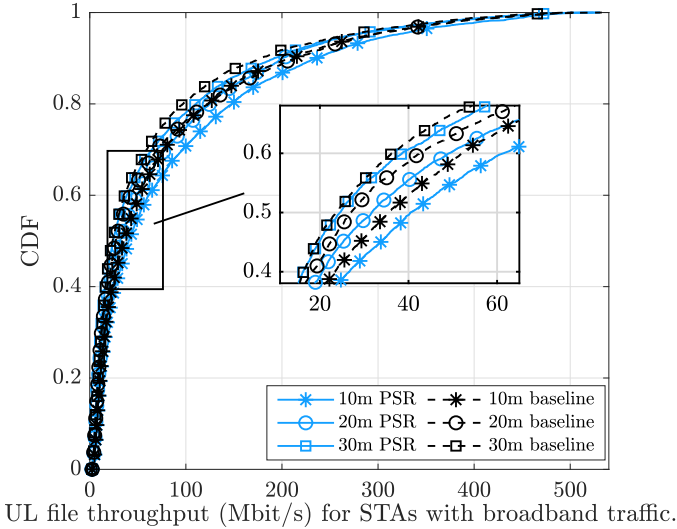}
\caption{CDF of the uplink file throughput for broadband STAs, inter-AP distances $d = \{ 10, 20, 30\}$m, and 100  Mbps broadband trafic load.}
\label{fig:Throughput}
\end{figure}

Fig.\ \ref{fig:Throughput} shows the impact of PSR on the throughput of the broadband stations. Overall, it can be observed that broadband stations also benefit from the implementation of the PSR feature. This is a direct consequence of the reduced number of devices that contend simultaneously for the wireless medium, and the resultant reduced channel access delays. The rationale for this behavior is illustrated in Fig. \ref{fig:psr}(b).

\subsection{Traffic Load}

\label{4b}

\begin{figure}[!t]
\centering
\includegraphics[width=\columnwidth]{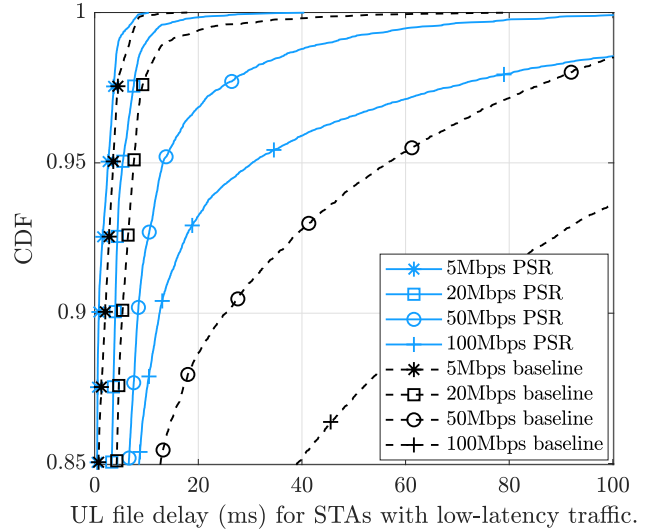}
\caption{CDF of the uplink file delay for low-latency STAs, traffic loads of \{5, 20, 50, 100\} Mbps, and inter-AP distance of 20 m.}
\label{fig:loaddelay}
\end{figure}

Fig.\ \ref{fig:loaddelay} quantifies the worst-case latency benefits of enabling PSR under different traffic loads. More concretely, Fig.\ \ref{fig:loaddelay} shows the CDF of the 15\% worst-case uplink file delays for low-latency STAs when the application layer traffic offered for broadband stations varies from 5  Mbps to 100   Mbps in the scenario with 20 m inter-AP distances. When comparing the behavior of the systems with and without PSR capabilities, two distinct regimes can be observed in Fig.\ \ref{fig:loaddelay}. First, for low and moderate traffic loads, i.e., up to 20   Mbps, both systems' latency performance is similar. This is because the wireless medium is generally available for transmission when devices transmit sporadically, which entails that devices do not need to contend for channel access for long periods. Second, for relatively high traffic loads, i.e., 50  Mbps and above, the system with PSR capabilities provides substantial latency benefits. Specifically, the implementation of PSR allows reducing the $15\%$ worst-case latencies by a factor of $\approx4.6\times$ when the offered traffic load reaches 100   Mbps. It is for these high traffic loads---where the wireless medium remains generally occupied---that allowing a more aggressive spatial reuse can provide remarkable latency benefits.

\subsection{Number of Antennas at the AP}

\label{4c}

\begin{figure}[!t]
\centering
\includegraphics[width=\columnwidth]{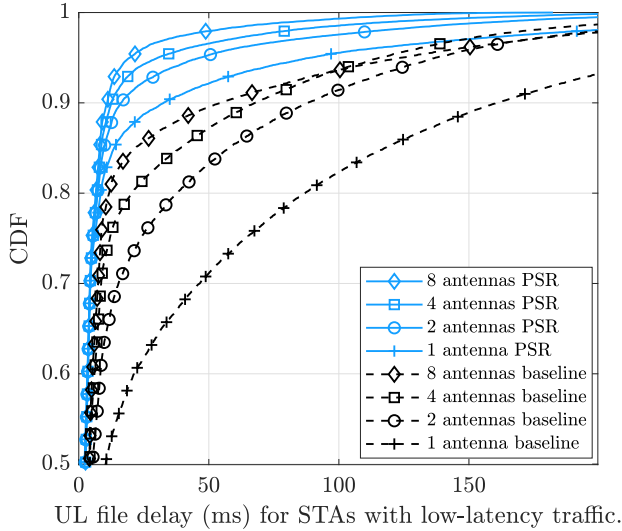}
\caption{CDF of the uplink file delay for low-latency STAs, inter-AP distance $d=$20 m, traffic load of 100  Mbps, and various numbers of AP antennas.}
\label{fig:antennadelay}
\end{figure}

Fig.\ \ref{fig:antennadelay} shows the upper part of the CDF of the uplink file delay---which includes the median and worst-case latencies---for the low-latency stations when equipping APs with 1, 2, 4 or 8 antennas. We consider a scenario with an inter-AP distance of 20 m and where the broadband stations offer a traffic load of 100  Mbps to stress the system load. As one could expect, the file latencies of 802.11ax systems with and without PSR capabilities are generally reduced when the number of antennas per AP grows.\footnote{ The crossover between the four- and eight-antenna scenarios is due to the system scheduling as many STAs as possible. This is, in general, suboptimal,  e.g., for APs with many antennas and spatial channel correlation among STAs.} We can also observe how, for a fixed number of AP antennas, a  PSR-capable system always outperforms the corresponding baseline system.

Remarkably, the results of Fig.\ \ref{fig:antennadelay} demonstrate that for a large portion of the CDF---up to the 95th percentile---a PSR-system with single-antenna APs can outperform a system without spatial reuse capabilities where the APs are equipped with eight antennas. In other words, allowing low-latency STAs to access the channel more aggressively is generally more effective than increasing the maximum AP throughput per transmission through multi-antenna capabilities for the scenarios under evaluation. Indeed, we can observe that the file latencies of PSR-capable systems are almost identical up to the 80th percentile of the CDF, regardless of the number of antennas per AP. Altogether, these results convey the important message that---for short-packet low-latency applications---the deployment of APs with PSR signal processing capabilities may be an appealing alternative to the deployment of the traditionally more costly multi-antenna APs.

\section{Conclusions}
\label{sec:conclusions}

In this paper, we have provided an extensive analysis of the benefits of enabling the PSR feature in 802.11ax WLANs in terms of reduced latency. Under realistic models, we evaluated the performance of PSR  for different scenario densities, traffic loads, and AP antenna arrays. Our studies show that, for  high-, medium-, and low-density deployments with high traffic loads, PSR can reduce worst-case latencies with respect to baseline 802.11ax operations by a factor 1.8, 3.8, and 3.4, respectively. The results also show that allowing a more aggressive spatial reuse in high-loaded scenarios provides the most remarkable latency benefits. Finally, we found that, for the scenarios under considerations, even single-antenna APs with PSR capabilities may outperform baseline 802.11ax multi-antenna APs, designating the PSR feature as an alternative to more costly multi-antenna APs for low-latency applications. Possible future work includes investigating the performance of PSR when combined with novel IEEE 802.11be features to further increase spatial reuse and reduce latency, such as coordinated beamforming and null steering \cite{psrContribution,lopez2019ieee,GarGerGal2018}.

\ifCLASSOPTIONcaptionsoff
  \newpage
\fi
\bibliographystyle{IEEEtran}
\bibliography{Strings_Gio,Bib_Gio}
\end{document}